\documentclass[runningheads]{llncs}
\usepackage{graphicx}
\usepackage{amsmath}
\usepackage{enumitem,amssymb}
\usepackage{booktabs} %
\usepackage{array}
\usepackage{tabularx}
\usepackage{seqsplit}
\usepackage{makecell}
\usepackage{tabularray}
\usepackage{hyphenat}
\usepackage{hyperref}
\usepackage{mathtools}
\usepackage{cite}

\usepackage{colortbl}
\newcommand{\blue}[1]{\textcolor{black}{#1}}

\begin{document}
\title{Answer Retrieval in \\Legal Community Question Answering}
\author{Arian Askari\orcidID{0000-0003-4712-832X} \and Zihui Yang \and Zhaochun Ren\orcidID{0000-0002-9076-6565} \and Suzan Verberne\orcidID{0000-0002-9609-9505}}
\authorrunning{Askari et al.}
\institute{Leiden University, Leiden, the Netherlands \email{a.askari@liacs.leidenuniv.nl}, \email{zoeyangzihui@gmail.com}, \email{s.verberne@liacs.leidenuniv.nl}, \email{z.ren@liacs.leidenuniv.nl} 
}
\maketitle              %
\begin{abstract}
The task of answer retrieval in the legal domain aims to help users to seek relevant legal advice from massive amounts of professional responses. 
Two main challenges hinder applying existing answer retrieval approaches in other domains to the legal domain: (1) a huge knowledge gap between lawyers and non-professionals; and (2) a mix of informal and formal content on legal QA websites.
To tackle these challenges, we propose CE\textsubscript{FS}, a novel cross-encoder (CE) re-ranker based on the \textbf{f}ine-grained \textbf{s}tructured inputs. CE\textsubscript{FS} uses additional structured information in the CQA data to improve the effectiveness of cross-encoder re-rankers.
Furthermore, we propose \textit{LegalQA}: a real-world benchmark dataset for evaluating answer retrieval in the legal domain.
Experiments conducted on LegalQA show that our proposed method significantly outperforms strong cross-encoder re-rankers fine-tuned on MS MARCO.
Our novel finding is that adding the question tags of each question besides the question description and title into the input of cross-encoder re-rankers structurally boosts the rankers' effectiveness.
While we study our proposed method in the legal domain, we believe that our method can be applied in similar applications in other domains.
\keywords{Legal Answer Retrieval \and Legal IR \and Data collection \and Fine-grained structured cross-encoder}
\end{abstract}
\section{\blue{Introduction}}%
As an established problem in information retrieval, answer\footnote{We interchangeably use the word answer and response to refer to the content written by the professional lawyer.} retrieval~\cite{bian2008finding} has been studied in a variety of domains, including healthcare~\cite{budler2023review}, social media~\cite{bian2008finding,xiong2019tweetqa}, and programming~\cite{yen2013support}.
Community question answering (CQA) platforms provide common sources for answer retrieval~\cite{roy2018finding,atkinson2013evolutionary,roy2023analysis}, e.g., Stackoverflow\footnote{https://stackoverflow.com/} for finding answers of programming questions~\cite{atkinson2013evolutionary}.
In the legal domain, users\footnote{We refer to the person who posts a question as a user or questioner throughout this paper.} seek legal advice from professionals, and timely access to accurate answers is important.
Legal language, often considered a sub-language due to its distinct characteristics, emphasizes the importance of answer retrieval in the legal domain~\cite{haigh2018legal,williams2007tradition,tiersma1999legal}.
However, answer retrieval has not been addressed in the legal domain yet.
\par
In this paper, we propose the task of legal answer retrieval. We start by analyzing the effectiveness of the widely used two-stage ranking pipeline~\cite{injectbm25}, which consists of an efficient retriever, e.g., BM25~\cite{robertson1994some}, to retrieve a shortlist of documents from the collection, and a re-ranker~\cite{nogueira2019passage} to increase the effectiveness of the initial ranking. On this premise, we propose cross-encoder\textsubscript{CAT} (CE\textsubscript{CAT}), which uses cross-encoders for optimizing the re-ranking stage by concatenating the query and the candidate document in the input~\cite{injectbm25}.
Given a new question and the corresponding initial ranked list produced by BM25, the CE\textsubscript{CAT} aims to effectively reorder the candidate documents from the initial ranked list to locate the most relevant responses on top of the ranked list. We fine-tune a cross-encoder re-ranker, called CE\textsubscript{FS}, based on our novel \textbf{f}ine-grained \textbf{s}tructured inputs to learn the relevance between a pair of a question and its best answer based on structured information from the data; such a model would be able to effectively re-rank prior existing relevant responses where the best answer is not provided \cite{yang2019end}.
\par
Our proposed method is inspired by recent studies that have shown task-level input modification could improve the effectiveness of cross-encoder re-rankers. For instance, BERT-FP~\cite{han2021fine} has shown that adding splitter tokens between each utterance of a dialogue can improve the effectiveness of BERT-based re-rankers. %
We are also motivated by the fact that, in legal CQA, question tags consist of important legal terms related to the legal question and using them can potentially bridge the knowledge gap between lawyer content and questioner content. It is noteworthy to mention that question tags are dynamically generated by the online platform that we have used in this paper.
\par
For the first attempt, we release a new benchmark dataset, namely LegalQA, in this paper. Each question in LegalQA has a corresponding best answer selected by the questioner \footnote{\href{https://github.com/arian-askari/AnswerRetrieval-Legal}{https://github.com/arian-askari/AnswerRetrieval-Legal}}. 
It consists of 9,846 questions and 33,670 answers responses by identified lawyers, organized in train, validation, and test splits. 
Our experiments on LegalQA show that CE\textsubscript{FS} significantly outperforms regular and strong fine-tuned cross-encoder re-rankers such as MiniLM-MSMARCO \cite{minilm_msmarco} which is a highly effective cross-encoder re-ranker trained on MS MARCO \cite{nguyen2016ms}. %
\par
Our contributions are as follows: (1) We formulate the task of answer retrieval in the legal domain and release LegalQA. As far as we know, this is the first benchmark test collection for legal answer retrieval.
(2) We evaluate the applicability of both probabilistic and existing effective fine-tuned cross-encoder re-rankers on legal answer retrieval. (3) We propose CE\textsubscript{FS} taking into account the different elements of a legal question with a fine-grained structured input that significantly outperforms regular fine-tuning cross-encoder re-rankers and the strong cross-encoder re-ranker trained on MS MARCO.

\section{\blue{Related work}}%
\blue{The objective of answer retrieval is to find the most relevant response given a question, which aligns with the core retrieval objective \cite{yang2019end,roy2018finding,atkinson2013evolutionary,roy2023analysis,bian2008finding,xiong2019tweetqa,yen2013support}.
Therefore, the effective CE\textsubscript{CAT} re-rankers are suitable to be invested in this task as they have shown high effectiveness in addressing various core retrieval tasks \cite{nogueira2019passage,abolghasemi2022improving,hofstatter2020improving,rau2022role}. 
Several studies have shown the impact of modifying the regular input of CE\textsubscript{CAT} could improve their effectiveness such as \cite{han2021fine,askari2023closercikm,injectbm25,boualili2020markedbert}.
However, there are no studies that investigated the usage of question tags to improve cross-encoder re-rankers in community question-answering systems.}
\blue{The most relevant work to this study is the recent work by \cite{mansouri2023falqu} that investigates the effectiveness of existing methods on legal CQA data.
However, the responses in that work are written by legal forum users rather than identified lawyers, and they do not improve the effectiveness of existing methods for this particular domain. %
They also included the fine-tuned MiniLM on MS Marco \cite{minilm_msmarco} in their study and showed its poor performance in the legal domain. 
Furthermore, Martinez-Gil et al. \cite{martinez2023survey} explore potential future directions for legal answering systems in a survey on legal systems designed for lawyers or law students including statute law retrieval and task legal textual entailment tasks.
In contrast, our focus is on answer retrieval in a legal community question-answering system with a combination of legal language by lawyers and everyday language by questioners who ask questions.}
\begin{figure}[t]
 \centering
 {\includegraphics[width=0.80\linewidth]{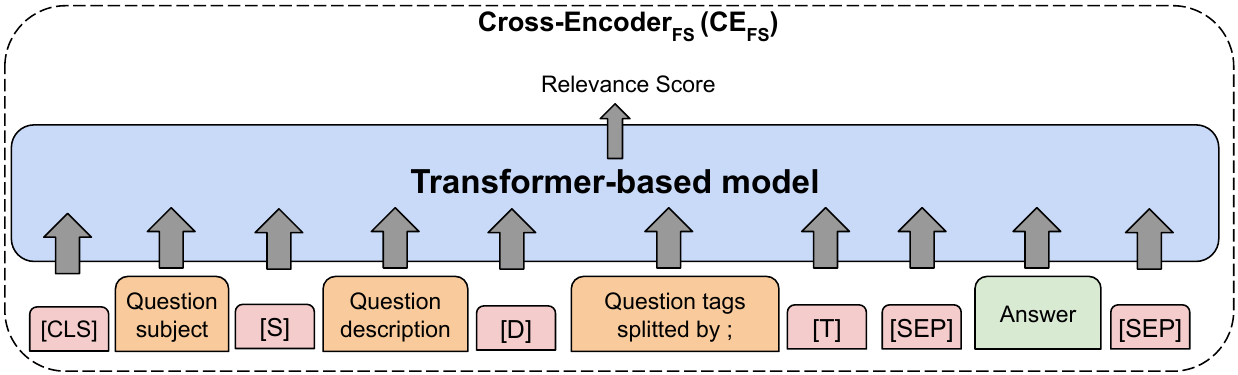}}
 \caption{Fine-grained structured input of CE\textsubscript{FS}}\label{fig:ce_fs}
\end{figure}
\section{Methods}
\blue{We present our method, CE\textsubscript{FS}, as an effective cross-encoder based re-ranker for the question answering retrieval in the legal domain. Our retrieval pipeline consists of first-stage retrieval and re-ranking.
For a user's questions, the first-stage retrieval returns a set of initial responses from the dataset. These results are then re-ranked in the second step to improve the effectiveness of retrieval.}
\par
\textbf{Collected dataset for the task.}
\blue{We create the LegalQA dataset by using the question URLs shared by \cite{AskariECIR22} to investigate expert finding in the legal domain. In the context of the expert finding task, the input consists solely of specific question tags, and the output is a list of experts about that specific question tag. In contrast, in answer retrieval, the standard input is the question's content, and the desired output is the most suitable answer to that question. We therefore create a new dataset, LegalQA based on the expert finding data. %
The LegalQA dataset is derived from a subset of the `bankruptcy' related forum on Avvo\footnote{\url{https://www.avvo.com/topics/bankruptcy}} in California, spanning from January 2008 to July 2021, and comprises 9,846 questions by regular users and 33,670 answers by certified lawyers. %
Notably, lawyers' profiles on Avvo are associated with their real names, distinguishing them from regular users. The questions are categorized, and each category, such as `bankruptcy,' includes questions with different category tags, for instance, `bankruptcy homestead exemption'.
Following \cite{AskariECIR22}, we determine a question's best answer based on whether it is selected as the most helpful by the questioner or if it receives ``lawyer agree'' votes from at least three certified lawyers, which is distinct from the ``helpful'' upvotes provided by other users.
To facilitate model training and evaluation, we split the dataset into three subsets: 70\% for training, 10\% for validation, and 20\% for testing, based on the chronological order of questions.
To ensure the dataset's integrity and eliminate duplicates, we perform a lexical similarity analysis using Levenshtein distance \cite{levenshtein_distance} on all the questions\footnote{We use the implementation available at \url{https://github.com/seatgeek/thefuzz}}. This analysis identified 50 pairs of questions with more than a 90\% overlap. In cases of high overlap between question pairs, we retain the longer question and discard the shorter one. Additionally, we reassign the responses from the removed question to the retained one.}
\par
\textbf{Baselines.}
\blue{For the first-stage retrieval, we employ BM25 as the baseline method. Additionally, we compare BM25 to a lexical-based probabilistic first-stage retriever named LMD\cite{ponte2017language} to assess the effectiveness of BM25 compared to another lexical-based retrieval. We leave out investigating the effectiveness of Transformer-based first-stage retrievers, e.g., dense retrievers, since we focus on improving the effectiveness of the re-ranking stage of answer retrieval in the legal domain. We leave further analysis on Transformer-based first-stage retrievers to future work.
For the re-ranking, we use cross-encoder re-rankers (\textsubscript{CE\textsubscript{CAT}}) in different settings: (1) a fine-tuned cross-encoder on MS MARCO; (2) a fine-tuned cross-encoder on the LegalQA training set; (3) a pre-trained cross-encoder as the zero-shot baseline.}%
\par
\blue{\textit{BM25 and LMD.} We use BM25 as first stage retriever, a commonly used ranking function that efficiently retrieves a set of documents from the full document collection based on word overlap \cite{BM25,Out-of-Domain_Semantics_to_the_Rescue_Zero-Shot_Hybrid_Retrieval_Models}.} %
\par
\blue{\textit{CE\textsubscript{CAT}. }We use CE\textsubscript{CAT} as a strong re-ranker. 
The query q1:m and answer content a1:n sequences are concatenated with the [SEP] token, and the [CLS] token representation computed by CE is scored with a single linear layer W in the CECAT ranking model: $CE_{CAT}(q_{1:m},a_{1:n}) = CE([CLS]\,q\,[SEP]\,a\,[SEP]) * W$.
We evaluate fine-tuned both MS MARCO-trained CE\textsubscript{CAT}, MiniLM-MSMARCO\cite{minilm_msmarco}, and LegalQA-trained CE\textsubscript{CAT} following by the above design. We use MiniLM \cite{wang2020minilm} as the cross-encoder model thorough all of the experiments.}
\par
\textbf{Proposed method: \blue{Cross-encoder\textsubscript{FS} (CE\textsubscript{FS})}.}
\blue{We propose CE\textsubscript{FS} in order to capture the relevance within the question and best answer based on a fine-grained structural-based input representation tailored for cross-encoder re-rankers.
In Figure \ref{fig:ce_fs}, we present the input representation of CE\textsubscript{FS}, which is formally explained as follows:
\begin{equation}\label{eq:SI_input_format}
    \resizebox{0.91\hsize}{!}{$CE_{FS}(q_{1:m},a_{1:n}) = CE([CLS]\,q_{Subject}\,[S]\,q_{Description}\,[D]\,q_{Tags}\,[T]\,[SEP]\,a\,[SEP]) * W$}
\end{equation}
Here, the [S], [D], and [T] tokens serve as separators, i.e., splitter tokens, for different parts of the question, namely Subject, Description, and Category tags, respectively. The input representation of CE\textsubscript{FS} is designed to take into account different aspects of the retrieval context. The novelties brought by CE\textsubscript{FS} can be summarized in two key aspects:}
\blue{\begin{itemize}
	\item \textit{Structured Input}. CE\textsubscript{FS} employs structured input by dividing the question into distinct sections – the subject, description, and tags. These sections are separated by splitter tokens. Such structuring not only facilitates a more comprehensive representation of the query but also emphasizes the importance of individual sections to the re-ranker since the cross-encoder in CE\textsubscript{FS} can comprehend different aspects of the information and assign varying levels of importance to each section of the question. 
	\item  \textit{Question Tags}. CE\textsubscript{FS} takes the question tags into account in a straightforward yet effective manner by incorporating them into the cross-encoder re-ranker's input. Each question tag is separated by a semicolon. The motivation behind these additions is to equip the re-ranker with more comprehensive knowledge about the query, enabling it to grasp both an overview of the legal question's topic and its detailed category tags.
\end{itemize}}

\section{Experiments and results}\label{sec:result}
\begin{table}[t]
\centering
\caption{Effectiveness results. $\dagger$ denotes a statistically significant improvement of \textbf{CE\textsubscript{FS}} over the second most effective re-ranker, CE\textsubscript{CAT} fine-tuned on the LegalQA training set. Statistical significance was measured with a paired t-test ($p < 0.001$) with Bonferroni correction for multiple testing. All re-rankers used top-1000 retrieved answers by BM25 as the initial ranking.}
\label{tab:effectiveness}
\resizebox{10cm}{!}{
    \begin{tabular}{l|l|c|c|c|c|c|c} 
    \toprule
    Training source      & $\,$Model$\,$     & $\,$MAP@1k$\,$ &  $\,$R@1k$\,$ & $\,$R@100$\,$ & $\,$R@10$\,$ & $\,$R@2$\,$ & $\,$R@1$\,$  \\ \midrule
    \multicolumn{8}{l}{\textbf{First-stage retrievers}}              \\ 
    \multicolumn{1}{c|}{---}             & $\,$BM25$\,$      & .120      & \textbf{.542}   & .354   & .192    & .113     & .069       \\
    \multicolumn{1}{c|}{---}             & $\,$LMD$\,$      & .080      & .540   & .321   & .153    & .840     & .050       \\ \midrule
    \multicolumn{8}{l}{\textbf{Cross-encoder re-rankers}}              \\ 
    MS MARCO       & $\,$CE\textsubscript{CAT}$\,$    & .109      & \textbf{.542}   & .341   & .173    & .101     & .087       \\ 
    LegalQA (\textbf{ours}) & $\,$CE\textsubscript{CAT}$\,$    & .236      & \textbf{.542}   & .495   & .381    & .204     & .181       \\
    LegalQA (\textbf{ours}) & $\,$CE\textsubscript{\textbf{FS}} \textbf{(ours) }     & \textbf{.270}$\dagger$      & \textbf{.542}   & \textbf{.524}$\dagger$   & \textbf{.428}$\dagger$    & \textbf{.261}$\dagger$     & \textbf{.209}$\dagger$       \\
    \bottomrule
    \end{tabular}
}
\end{table}
\blue{\textbf{Experimental setup.} We use the Huggingface library \cite{wolf2019huggingface} for the cross-encoder re-ranking training and inference. We add the splitter tokens into the tokenizer of the cross-encoder. Following prior work \cite{hofstatter2020improving} we use Cross-Entropy loss \cite{zhang2018generalized}, training batch size of 32, and Adam \cite{kingma2014adam} optimizer with a learning rate of $7*10^{-6}$ for all cross-encoder layers, regardless of the number of layers trained.}%
\par
\blue{\textbf{Ranking quality.} Table \ref{tab:effectiveness} illustrates the effectiveness of lexical-based first-stage retrievers and cross-encoder re-rankers.
For re-ranking, our proposed method, CE\textsubscript{FS}, demonstrates significant improvements over both CE\textsubscript{CAT} fine-tuned on the LegalQA dataset and CE\textsubscript{CAT} trained on MS MARCO, referred to as MiniLM-MSMARCO \cite{minilm_msmarco}.  E.g., in terms of MAP@1k, CE\textsubscript{FS} achieves 0.270 vs. 0.236 for CE\textsubscript{CAT} and 0.109 for MiniLM-MSMARCO. The higher effectiveness of CE\textsubscript{FS} over the CE\textsubscript{CAT} confirms the effectiveness of the proposed method, and the low performance of MiniLM-MSMARCO on LegalQA confirms the challenges of the legal domain since MiniLM-MSMARCO achieves a three times higher effectiveness on MS MARCO dataset in terms of MAP@1k. Among the first-stage retrievers, BM25 outperforms LMD in terms of initial ranking effectiveness. The relatively low effectiveness of BM25 reveals a noticeable difference in lexical word overlap between the question and the best answer. This difference serves as an indicator of a knowledge gap between the questioner and the lawyer, resulting in a higher occurrence of lexical mismatches between the question and the relevant answer. Achieving a higher overall effectiveness in this task is a potential area for improvement in future works.}
\par
\begin{table}[t]
\centering
\caption{Results of the ablation study on CE\textsubscript{FS}.}
\label{tab:ablation}
\resizebox{10cm}{!}{
    \begin{tabular}{l|c|c|c|c}
    \toprule
    Model                                                & MAP@1k & R@100 & R@10 & R@1   \\ \midrule
    CE\textsubscript{FS} w/o [T] splitter and query tags           & .252   & .510  & .405 & .163  \\ %
    CE\textsubscript{FS} w/o [S] splitter and question subject     & .239   & .498  & .379 & .154  \\   %
    CE\textsubscript{FS} w/o [D] splitter and question description & .191   & .461  & .315 & .119  \\ \midrule%
    CE\textsubscript{FS}                                          & \textbf{.270}   & \textbf{.524}  & \textbf{.428} & \textbf{.209}  \\
    \bottomrule
    \end{tabular}
}
\end{table}
\blue{\textbf{Ablation study.} We do an ablation study on the CE\textsubscript{FS} to analyze to what extent each section of the fine-grained structured input of CE\textsubscript{FS} has an impact on the effectiveness of CE\textsubscript{FS}. As shown in Table \ref{tab:ablation}, the effectiveness of CE\textsubscript{FS} is highest when we use all of the splitter and corresponding contents. We see that query tags and [T] splittor have less impact on the effectiveness and question description and [D] has the highest impact. Question subject and [S] have the second-highest impact. This suggests that although the query tags have a role in improved effectiveness, question subject and description have still a large impact on the effectiveness.}
\section{Conclusion}
\blue{For investigating answer retrieval in legal community question answering, we created a dedicated dataset, called LegelQA, which we divide into training, validation, and test sets. We use this dataset to train neural retrieval models, introducing a novel and highly effective re-ranker called CE\textsubscript{FS}, and take a fine-grained structured approach to leverage the information available in the legal domain, demonstrating significantly higher effectiveness over common strong cross-encoder re-rankers. %
We investigate the impact of each part of the fine-grained structured input within our method and highlight the significant role played by question tags in improving retrieval effectiveness besides the most important part of the question which is question description. While our method is initially proposed for answer retrieval within the legal domain, we foresee its potential application in answer retrieval for community question-answering systems across various domains where question tags are provided alongside each question post. For future work, our data, LegalQA, facilitates other tasks such as legal response generation.}
\section*{ACKNOWLEDGMENTS}
This work was supported by the EU Horizon 2020 ITN/ETN on Domain Specific Systems for Information Extraction and Retrieval (H2020-EU.1.3.1., ID: 860721).
\bibliographystyle{splncs04}
\bibliography{main.bib}
\end{document}